\documentclass[mypaper,8pt,twoside]{CoAst}
\usepackage{epsf,graphicx,fancyhdr}
\usepackage{sfmath}

\usepackage{amsmath}

\renewcommand{\chaptermark}[1]{\markboth{#1}{}}

\newcommand{\kopf}{\small\itshape Comm. in Asteroseismology\\ Vol. 150, 2007}
\newcommand{\Authors}[1]{\begin{center}\normalsize\bf\sf #1 \end{center}}

\renewcommand{\author}[1]{\begin{center}\normalsize\bf\sf #1 \end{center}}
\newcommand{\Address}[1]{\begin{center}\small\sf #1 \end{center}}

\newcommand{\References}[1]{\vspace{2.4mm}\begin{flushleft}{\large References\\}\vspace*{1mm}\small #1 \end{flushleft}}

\newcommand{\chapterDSSN}[2]{\chapter[\sf\normalsize #1\\ \footnotesize \hspace*{5mm}by #2 \sf\normalsize][]{#1\\}\rhead[\fancyplain{}{\sf\footnotesize \center{#1}}]{\fancyplain{}{\sffamily\thepage}}\lhead[\fancyplain{\kopf}{\sffamily\thepage}]{\fancyplain{\kopf}{\sf\footnotesize \center{#2}}}}

\newcommand{\figureDSSN}[5]{\begin{figure}[#4]
\centering
\includegraphics*[#5]{#1}
\caption{#2}
\label{#3}
\end{figure}}

\renewcommand{\chaptername}{}
\newcommand{\acknowledgments}[1]{\vspace*{5mm}\noindent\begin{bf}Acknowledgments. \end{bf} #1}

\def\rfr{\smallskip\par\noindent
        \hangindent=7truemm
        \hangafter=1}

\begin{document}
\sf

\chapterDSSN{The Beta Cephei instability domain for the new solar 
composition and with new OP opacities}{A. A. Pamyatnykh and W. Ziomek}

\Authors{A. A. Pamyatnykh,$^{1,2,3}$ W. ZiomekÂ\,$^4$}
\Address{
$^1$\,Copernicus Astronomical Center, Bartycka 18, 00-716 Warsaw, Poland\\
$^2$\,Institute of Astronomy, Pyatnitskaya Str. 48, 109017 Moscow, 
Russia\\
$^3$\,Institut f\"ur Astronomie, T\"urkenschanzstrasse 17, 1180 Vienna, 
Austria\\
$^4$\,Astronomical Institute, Kopernika 11, 51-622 Wroclaw, Poland
}

\noindent
The recent revision of the solar chemical composition (A04: Asplund et
al.\ 2005) leads to a decrease of about 40\% in the C, N, O, Ne
abundances and to a $\sim$ 20~\% decrease of Fe and some other metal
abundances in comparison with older abundances (GN93: Grevesse \& Noels
1993), as shown in Fig.\,1.

\figureDSSN {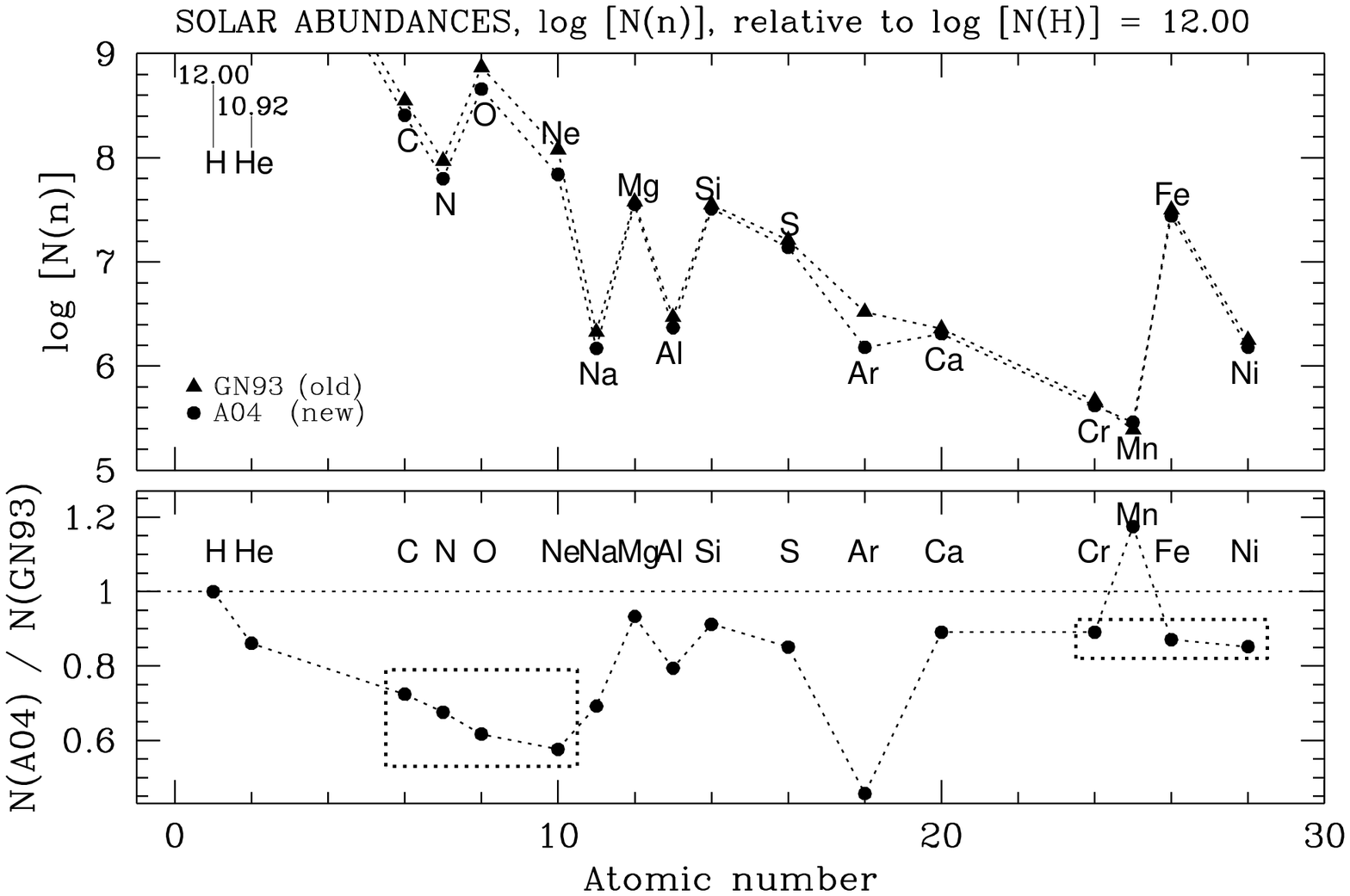} {The new solar abundances in comparison
with the older ones.} {freqspec}{h}{width=113mm}

\noindent We tested the effects of these modifications of the heavy
element abundances on the instability of $\beta$ Cephei models. For
opacities, the newest data from the Opacity Project (Seaton 2005) were
used. Fig.\,2 shows that the $\beta$~Cephei instability domain in the HRD,
when computed with new data for $Z=0.012$ (revised solar value), is very
similar to the instability domain computed with the OPAL opacities
(Iglesias \& Rogers 1996) for older solar metallicities and $Z=0.02$. For
the older data and assuming $Z=0.012$, we obtain only weak $\beta$~Cep
instability (Pamyatnykh 1999). Two effects are responsible for stronger
instability when using the new data: (i) The metal opacity bump in the OP
case is located slightly deeper in the star than that in the OPAL case,
which results in more effective driving; (ii) at a fixed $Z$ value, the
new Fe-group abundances are higher than the older ones because the $Z$
value is determined mainly by the abundances of C, N, O, and Ne (see
Fig.\,1).

\figureDSSN{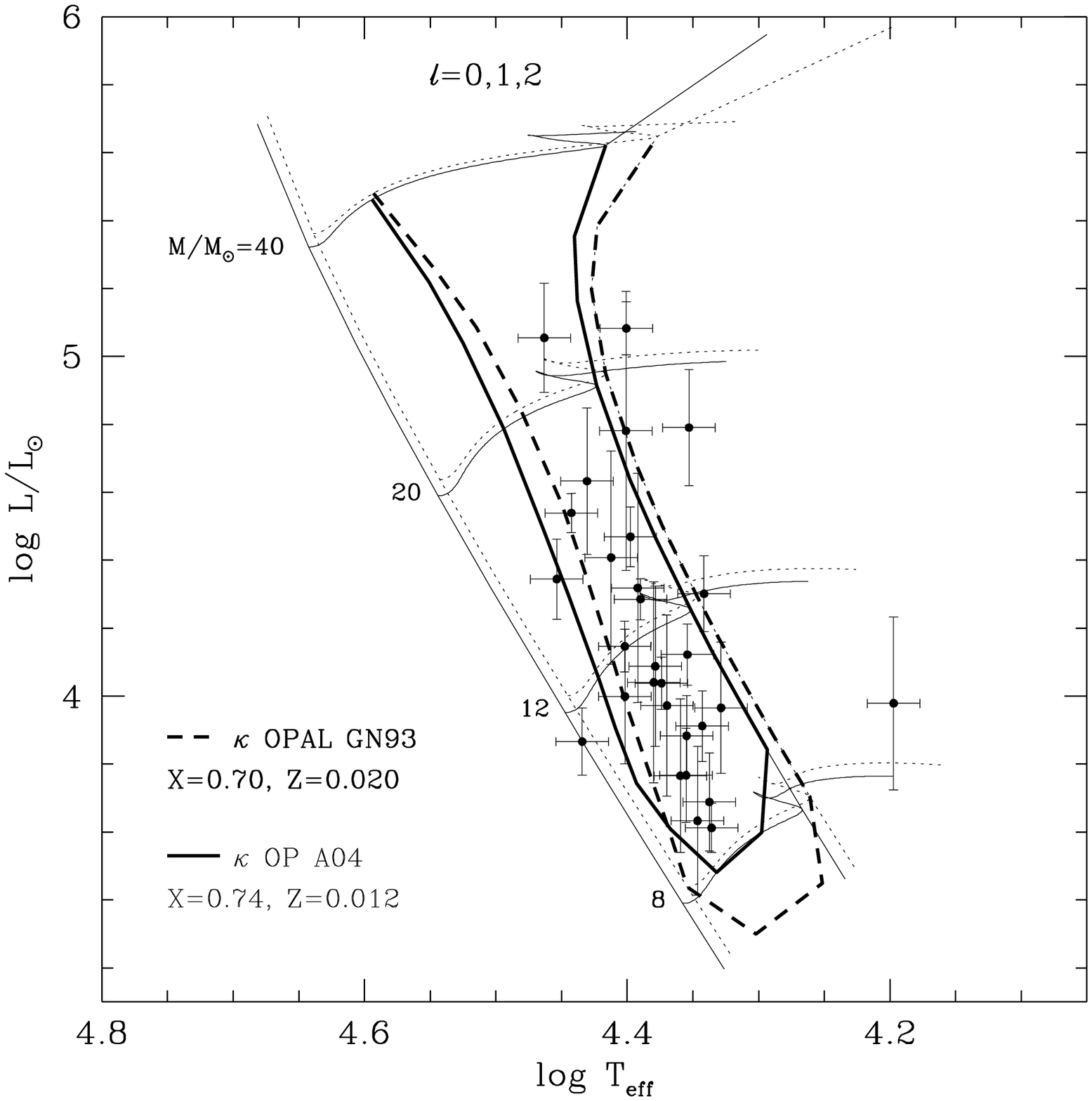}{The new $\beta$~Cephei instability
domain in the main-sequence band (OP opacity, A04 mixture, $Z=0.012$)
compared with the older one (OPAL GN93, $Z=0.02$, see Pamyatnykh 1999). 29
bright variables from Stankov \& Handler (2005) with $m_{_{\rm V}}<6.0$
and well-measured Hipparcos parallaxes are
plotted.}{freqspec2}{ht}{width=102mm}

\acknowledgments{AAP acknowledges partial financial support from
the HELAS project and from the Polish MNiI grant No.\ 1 P03D 021
28.}

\References{
\rfr Asplund M., Grevesse N., Sauval A.\ J., 2005, in Barnes III T.\ G., 
     Bash F.\ N., eds, ASP Conf. Ser. Vol.\,336, The Solar Chemical 
     Composition. Astron. Soc. Pac., San Francisco, p.\,25
\rfr Grevesse N., Noels A., 1993, in Pratzo N., Vangioni-Flam E., 
     Casse M., eds, Origin and Evolution of the Elements. Cambridge 
     Univ. Press, Cambridge, p.\,15
\rfr Iglesias C.\ A., Rogers F.\ J., 1996, ApJ, 464, 943
\rfr Pamyatnykh A.\ A., 1999, Acta Astron., 49, 119
\rfr Seaton M.\ J., 2005, MNRAS, 362, L1
\rfr Stankov A., Handler G., 2005, ApJS, 158, 193
}
\end{document}